\documentclass[12pt]{article}
\usepackage[latin1]{inputenc}
\usepackage{amsmath}  
\usepackage{english} 

\usepackage{amssymb} 
\usepackage{verbatim}  

\usepackage{fancyheadings} 
\renewcommand{\subsectionmark}[1]{}
\font \tensans                = cmss10
\font \fivesans               = cmss10 at 5pt

\font \sevensans              = cmss10 at 7pt

\newfam\sansfam
\textfont\sansfam=\tensans\scriptfont\sansfam=\sevensans
\scriptscriptfont\sansfam=\fivesans

\def\sans{\fam\sansfam\tensans}
\def\wwwr{{\rm I\!R}} 

\def\wwwc{{\mathchoice {\setbox0=\hbox{$\displaystyle\rm C$}\hbox{\hbox
to0pt{\kern0.4\wd0\vrule height0.9\ht0\hss}\box0}}
{\setbox0=\hbox{$\textstyle\rm C$}\hbox{\hbox
to0pt{\kern0.4\wd0\vrule height0.9\ht0\hss}\box0}}
{\setbox0=\hbox{$\scriptstyle\rm C$}\hbox{\hbox
to0pt{\kern0.4\wd0\vrule height0.9\ht0\hss}\box0}}
{\setbox0=\hbox{$\scriptscriptstyle\rm C$}\hbox{\hbox
to0pt{\kern0.4\wd0\vrule height0.9\ht0\hss}\box0}}}}

\def\wwwq{{\mathchoice {\setbox0=\hbox{$\displaystyle\rm Q$}\hbox{\raise
0.15\ht0\hbox to0pt{\kern0.4\wd0\vrule height0.8\ht0\hss}\box0}}
{\setbox0=\hbox{$\textstyle\rm Q$}\hbox{\raise
0.15\ht0\hbox to0pt{\kern0.4\wd0\vrule height0.8\ht0\hss}\box0}}
{\setbox0=\hbox{$\scriptstyle\rm Q$}\hbox{\raise
0.15\ht0\hbox to0pt{\kern0.4\wd0\vrule height0.7\ht0\hss}\box0}}
{\setbox0=\hbox{$\scriptscriptstyle\rm Q$}\hbox{\raise
0.15\ht0\hbox to0pt{\kern0.4\wd0\vrule height0.7\ht0\hss}\box0}}}}

\def\wwwz{{\mathchoice {\hbox{$\sans\textstyle Z\kern-0.4em Z$}}
{\hbox{$\sans\textstyle Z\kern-0.4em Z$}}
{\hbox{$\sans\scriptstyle Z\kern-0.3em Z$}}
{\hbox{$\sans\scriptscriptstyle Z\kern-0.2em Z$}}}}

\def\wwnegeinr{\ \\[1ex]\hspace{1em}}
\def\dpast{\hat{\Delta}}

\begin{document}
\title{A discrete and finite approach to past proper time}

\author{Wolfgang Orthuber\\
\footnotesize Klinik für Kieferorthopädie der Universität Kiel \\*[-0.2cm]
\footnotesize Arnold-Heller-Str. 16, D-24105 Kiel, Germany \\*[-0.2cm]
\footnotesize Email: mail@orthuber.com \\*[-0.2cm]
\footnotesize further info: http://www.orthuber.com/   }

\date{}

\maketitle 

\begin{abstract}
\hspace{-2em}
The function $\gamma(x)=\frac{1}{\sqrt{1-x^2}}$ plays an important role in mathematical physics, e.g. as factor for relativistic time dilation in case of $x=\beta$ with $\beta=\frac{v}{c}$ or $\beta=\frac{pc}{E}$. Due to former considerations \cite{or1} it is reasonable to study the power series expansion of $\gamma(x)$. Here its relationship to the binomial distribution is shown, especially the fact, that the summands of the power series correspond to the return probabilities to the starting point (local coordinates, configuration or state) of a Bernoulli random walk. So $\gamma(x)$ and with that also proper time is proportional to the sum of the return probabilities. In case of $x=1$ or $v=c$ the random walk is symmetric. Random walks with \linebreak absorbing barriers are introduced in the appendix. Here essentially the basic mathematical facts are shown and references are given, most interpretation is left to the reader.\\
\end{abstract}

\newpage

\setcounter{tocdepth}{4}
\tableofcontents

\section{Introduction}
It has been shown \cite{or1}, that the (measurable) result data vector of a physical experiment (with finite duration) can be calculated from the (measurable) initial data vector by combining a \emph{finite} number of basic arithmetic operations (addition, subtraction, multiplication, division). This doesn't contradict the fact, that many analytical functions with \emph{infinite} power series expansions can successfully predict (approximative) experimental results: They are only successful in case of convergence, i.e. in case of convergence of the partial sum sequence of the corresponding power series expansion. These partial sums can be calculated by finitely many basic arithmetical operations and the result is arbitrarily near to the one of the function. So there always can be an exact partial sum and an approximative function result without chance for experimental distinction. However, the study of partial sums is one possibility to learn more about the nature of the underlying (finite) physical process - even in case of missing convergence.\\
Here we study the function
\begin {equation}
\label{eqdefgamma}
\gamma: \ ]-1,1 [\ \to \wwwr, \ \ 
\gamma (x)=\frac{1}{\sqrt{1-x^2}}
\end {equation}
which is frequently used in mathematical physics, e.g. as factor for relativistic time dilation in case of $x=\beta$ with $\beta=\frac{v}{c}$ or $\beta=\frac{pc}{E}$ . We investigate the power series representation of $\gamma(x)$ and show its (simple?) relationship to the binomial distribution, which plays an important role in nature, often in not obvious way \cite{firstbinref}-\cite{lastbinref}\footnote{Recall the close connection between relativistic mass increase and time dilation, especially when reading \cite{bottini}, in which concrete physical relevance of finite partial sums (of the power series expansion of $\frac{1}{\sqrt{1-\hat x}})$ is shown.}. In the appendix we look also at $\frac{1}{\gamma(x)}$.

\section{The connection of proper time and return probabilities}
\subsection{The binomial series}
In case of $\alpha\in\wwwz^*=\{j\in\wwwz|j \ge 0\}$ the function
$$
\hat f_\alpha: \wwwc \to \wwwc \ \ \hat f_\alpha(z)=(1+z)^\alpha 
$$
has a finite power series expansion of the form
$$
\hat f_\alpha(z)
=1+\binom{\alpha}{1}z^{l}+\binom{\alpha}{2}z^{2}+...+\binom{\alpha}{\alpha}z^{\alpha}
=\sum_{l=0}^{\alpha}\binom{\alpha}{l}z^{l},
$$
in which $\binom{\alpha}{l}$ are the \emph{binomial coefficients}. These are defined by
\begin{equation}
\label{eqdefbin}
\binom{\alpha}{0}=1 \text{\ \ \ and \ \ \  }
\binom{\alpha}{l}=\frac{\alpha(\alpha-1)(\alpha-2)...(\alpha-l+1)}{l!}\ \ 
\text{ for } l \in \wwwz^*\backslash\{0\}\ .
\end{equation}
In case of $\alpha\in\wwwc\backslash\wwwz^*$ and $|z|<1$ we can develop the function $\hat f_\alpha(z)=(1+z)^\alpha$ into a convergent MacLaurin series \cite{go} \cite{kn} \cite{mac} \cite{mar}. If $f_\alpha(z)$ denotes the principal value of $\hat f_\alpha(z)$, which is equal to one at $z=0$, we obtain
$$
f_\alpha(0)=1, \ \  f_\alpha'(0)=\alpha, \ \  f_\alpha''(0)=\alpha(\alpha-1),
\ ... , \ f_\alpha^{(l)}=\binom{\alpha}{l} l!
$$
from which follows the representation of $f_\alpha(z)$ as \emph{binomial series}
\begin{equation}
\label{eqbinseries}
f_\alpha(z)=\sum_{l=0}^{\infty}\binom{\alpha}{l}z^{l} \ .
\end{equation}

\subsection{The power series of $\gamma (x)=\frac{1}{\sqrt{1-x^2}}$}
\label{Chapterpowseriesgamma}
Because of $\frac{1}{\sqrt{1+z}}=f_{-1/2}(z)$ we get with (\ref {eqdefbin}) and (\ref {eqbinseries})
\footnotesize
\begin{eqnarray}
\frac{1}{ \sqrt{1+z}}
&=&\sum_{l=0}^{\infty}\binom{-\frac{1}{2}}{l}z^{l} \notag\\
&=&1+ \frac{- \frac{1}{2}}{1} z^1+ \frac{- \frac{1}{2} \cdot - \frac{3}{2}}{1 \cdot 2} z^2
+ \frac{- \frac{1}{2} \cdot - \frac{3}{2} \cdot - \frac{5}{2}}{1 \cdot 2 \cdot 3} z^3
+ \frac{- \frac{1}{2} \cdot - \frac{3}{2} \cdot - \frac{5}{2} \cdot - \frac{7}{2}}{1 \cdot 2 \cdot 3 \cdot 4} z^4+... \notag\\
&=&1- \frac{1}{2^1 \cdot 1!} z^1+ \frac{1 \cdot 3}{2^2 \cdot 2!} z^2
- \frac{1 \cdot 3 \cdot 5}{2^3 \cdot 3!} z^3
+ \frac{1 \cdot 3 \cdot 5 \cdot 7}{2^4 \cdot 4!} z^4-... \notag\\
&=&1- \frac{2!}{(2^1 \cdot 1!)^2} z^1+ \frac{4!}{(2^2 \cdot 2!)^2} z^2
- \frac{6!}{(2^3 \cdot 3!)^2} z^3
+ \frac{8!}{(2^4 \cdot 4!)^2} z^4-... \notag\\
&=&\sum_{l=0}^{\infty}(-1)^l\frac{(2l)!}{(2^l \cdot l!)^2}z^{l} \notag\\
\label{eqherlgamma}
&=&\sum_{l=0}^{\infty}\binom{2l}{l} \left( \frac{-z}{4} \right)^{l} 
\end{eqnarray}
\normalsize
and after substitution of $z$ by $-x^2$
\begin{equation}
\label{eqgammaseries}
\gamma(x)=\frac{1}{ \sqrt{1-x^2}}=\sum_{l=0}^{\infty}\binom{2l}{l} \left( \frac{x}{2} \right)^{2l} \ .
\end{equation}

\subsection{Bernoulli random walk}
\label{ChapterBernoulliRandomWalk} 
A Bernoulli random walk is a stochastic process generated by a sequence of Bernoulli trials\footnote{Independent trials, each one of which can have only two results, e.g. "positive" (with probability $p$) or "negative" (with probability $1-p$).}
\cite{feller}\cite{spi}. It can be interpreted as a model for the movement of a particle in a one-dimensional discrete state space and may be described in the following terms: The particle moves "randomly" along a line over a lattice of equidistant points ("states"), which are indexed by an integer coordinate $k$. With every trial the particle makes a step from point $k$ to point $k+1$ with given probability $p$ ("positive direction") or a step from point $k$ to point $k-1$ with probability $1-p$ ("negative direction").\\\\
For $n\in\{1, 2, 3,...\}$ let's denote by $Q0P(n,k,p)$ the probability, that the particle is at point $k$ after the $n$-th step and by $Q0P(0,k,p)$ this probability before the first step. We assume start of movement at $k=0$, so 
$Q0P(0,0,p)=1 \text{ and } Q0P(0,k,p)=0 \text{ for } k\ne 0 $\\
 and furthermore
\begin{equation}
\label{eqwalklaw}
Q0P(n + 1, k, p) = p \ Q0P(n, k - 1, p) + (1 - p) \  Q0P(n, k + 1, p) \ .
\end{equation}
When making $n$ trials, point $k$ is only within reach, if $n-k$ and $n+k$ are non-negative even numbers. We will presuppose this subsequently. There are exactly $\binom{n}{\frac{n+k}{2}}$ ways with $\frac{n+k}{2}$ steps in positive and $\frac{n-k}{2}$ steps in negative direction, which lead into point $k$ after the $n$-th step. They respectively have the probability $(1-p)^{(n-k)/2} \ p^{(n+k)/2}$. So
the chaining of these Bernoulli trials results into the binomial distribution
\begin{equation}
\label{eqdefq0p}
Q0P(n,k,p)=\binom{n}{\frac{n+k}{2}} \ p^{(n+k)/2}\  (1-p)^{(n-k)/2} \ .
\end{equation}
We now look at the probabilities of return to the starting point. Because the movement started at $k=0$ these correspond to
$$
Q0P(2n,0,p)=\binom{2n}{n} (1-p)^n \ p^n\ ,
$$
i.e. $Q0P(2n,0,p)$ is the \emph{return probability} after the $2n$-th step\footnote{Return is only possible after an even number of steps.}. Substitution of $p$ by $\frac{1-\sqrt{1-x^2}}{2}$ or $\frac{1+\sqrt{1-x^2}}{2}$ yields
\begin{eqnarray}
&&Q0P(2n,0,\frac{1-\sqrt{1-x^2}}{2})\notag\\
&=&Q0P(2n,0,\frac{1+\sqrt{1-x^2}}{2})\notag\\
&=&\binom{2n}{n} \left(\frac{1-\sqrt{1-x^2}}{2} \ \ \frac{1+\sqrt{1-x^2}}{2}\right)^n\notag\\
\label{Q0PsubstitutedP}
&=&\binom{2n}{n} \left(\frac{x^2}{4}\right)^n=\binom{2n}{n} \left(\frac{x}{2}\right)^{2n}
\end{eqnarray}
and with (\ref {eqgammaseries}) we obtain
\begin{eqnarray}
\gamma(x)=\frac{1}{ \sqrt{1-x^2}}&=&\sum_{n=0}^{\infty}Q0P(2n,0,\frac{1-\sqrt{1-x^2}}{2})\notag\\
&=&\sum_{n=0}^{\infty}Q0P(2n,0,\frac{1+\sqrt{1-x^2}}{2})\ .
\label{eqgammaprob}
\end{eqnarray}
Note, that the condition
$$
p\in\left\{\frac{1-\sqrt{1-x^2}}{2}, \frac{1+\sqrt{1-x^2}}{2} \right\}
$$
is equivalent to
\begin{equation}
\label{eqp1p}
4p(1-p)=x^2 \ .
\end{equation}

Before we continue, we should remember, that the function $\gamma(x)$ cannot have an exact equivalent in physical (past) reality, because the sum on the right side of (\ref{eqgammaprob}) is not finite. Furthermore the values $Q0P(2n,0,p)$ are probabilities, and every expectation value calculated from probabilities has only an average, approximative meaning. Therefore we presuppose, that the random walk contains a sufficiently large number of steps, so that there can be an equivalent to a finite partial sum of the right side of (\ref{eqgammaprob}) sufficiently close to $\gamma(x)$, that the reliability of the expectation value calculated from it is so great, that the difference between the individual (discrete\footnote{Due to \cite{or1} the set of all possible measurement results is finite and so of course also discrete.}) measurement result and the calculated value isn't significant. With that we can summarize:

\subsection{Theorem: Proper time proportional to the sum of return probabilities}
\label{theoremgamma}
Let $\gamma(x)=\frac{1}{\sqrt{1-x^2}}$ represent the (approximative) time dilation factor of reference system A relative\footnote{We can assume $x=\frac{v}{c}$ if B is moving with velocity $v$ relative to A and space-time is flat.} to reference system B. Then proper time of A relative to B is (approximatively) proportional to the sum of the return probabilities to the starting point of a Bernoulli random walk, in which each step is directed from point\footnote{Every point can represent a state in a one-dimensional discrete state space and $k$ the integer index to it. Reversal of the order of the index is possible and has the same effect as exchange of the probabilities $p$ and $1-p$\ .} $k$ to $k+1$ with probability $p$, from point $k$ to $k-1$ with probability $1-p$ and $4p(1-p)=x^2$.\\

\subsection{Case $x=1$ resp. $v=c$}
\label{ChapterSymmetricGamma}
In many physical situations $x=1$, especially if $x=\frac{v}{c}$ and $v=c$ is the velocity of light resp. photons\footnote{Because $v=c$ is the maximal speed of information transport, this case is also important from information theoretical point of view.}. So the case $x=1$ is extremely frequent. Why?\\

The above consideration (\ref{eqp1p}) shows, that $x=1$ corresponds to $p=1-p=\frac{1}{2}$, i.e. the probabilities $p$ and $1-p$ of positive and negative step direction are equal. Now the reason of $x=1$ resp. $v=c$ for photons becomes clear: At this both directions have the same chance. Nature \emph{a priori} makes no preferences.\\


\subsubsection{Symmetric random walk}
In case of $x=1$ because of $p=1-p=\frac{1}{2}$ the random walk is symmetric. The accompanying probabilities are
\begin{equation}
\label{eqdefq0}
Q0(n,k):=Q0P(n,k,\frac{1}{2})=\binom{n}{\frac{n+k}{2}} \left( \frac{1}{2} \right)^n   .
\end{equation}

\begin{table}[h]
$ $\\
\setlength{\tabcolsep}{0.14cm}
\begin{tabular}{rrrrrrrrrrrrrrrrrr}
$n$ &$k\to$ & $-6$ & $-5$ & $-4$ & $-3$ & $-2$ & $-1$ & 
$\ \ 0$ & $\ \ 1$ & $\ \ 2$ & $\ \ 3$ & $\ \ 4$ & $\ \ 5$ & $\ \ 6$ & &\\ 

$\downarrow$ &$  $ & $  $ & $  $ & $  $ & $  $ & $  $ & $  $ & 
$  $ &$  $ & $  $ & $  $ & $  $ & $  $ & $  $ & $  $ & $$\\

$ 0$ &$  $ & $  $ & $  $ & $  $ & $  $ & $  $ & $  $ & 
$ \underline{1}$ &$  $ & $  $ & $  $ & $  $ & $  $ & $  $ & $  $ & $\ \cdot2^{0\ \ }$\\

$ 1$ &$  $ & $  $ & $  $ & $  $ & $  $ & $  $ & $ 1$ & 
$  $ &$ 1$ & $  $ & $  $ & $  $ & $  $ & $  $ & $  $ & $\ \cdot 2^{-1}$\\

$ 2$ &$  $ & $  $ & $  $ & $  $ & $  $ & $ 1$ & $  $ & 
$ \underline{2}$ &$  $ & $ 1$ & $  $ & $  $ & $  $ & $  $ & $  $ & $\ \cdot 2^{-2}$\\

$ 3$ &$  $ & $  $ & $  $ & $  $ & $ 1$ & $  $ & $ 3$ & 
$  $ &$ 3$ & $  $ & $ 1$ & $  $ & $  $ & $  $ & $  $ & $\ \cdot 2^{-3}$\\

$ 4$ &$  $ & $  $ & $  $ & $ 1$ & $  $ & $ 4$ & $  $ & 
$ \underline{6}$ &$  $ & $ 4$ & $  $ & $ 1$ & $  $ & $  $ & $  $ & $\ \cdot 2^{-4}$\\

$ 5$ &$  $ & $  $ & $ 1$ & $  $ & $ 5$ & $  $ & $10$ & 
$  $ &$10$ & $  $ & $ 5$ & $  $ & $ 1$ & $  $ & $  $ & $\ \cdot 2^{-5}$\\

$ 6$ &$  $ & $ 1$ & $  $ & $ 6$ & $  $ & $15$ & $  $ & 
$\underline{20}$ &$  $ & $15$ & $  $ & $ 6$ & $  $ & $ 1$ & $  $ & $\ \cdot 2^{-6}$\\

$...$ &$  $ & $  $ & $  $ & $  $ & $  $ & $  $ & $  $ & 
$  $ &$  $ & $  $ & $  $ & $  $ & $  $ & $  $ & $  $ & $$
\end{tabular}
\caption{The first values of $Q0(n,k)$. The return probabilities are underlined. The representation is chosen in a way that the well known Pascal triangle gets visible. Also the small modification in comparison with usual tables of binomial coefficients gets clear: The definition of $Q0(n,k)$ takes into account the symmetry. The underlined return probabilities are located in the symmetry center and have all the same index $k=0$.}
\label{tableQ0}
\end{table}

\subsubsection{Finite random walk - finite partial sum $\gamma_{2n}(x)$ of $\gamma(x)$}
\label{ChapterGammaPartialsum}
We now look again to $\gamma(x)$. In case $x=1$ the series (\ref{eqgammaprob}) doesn't converge, i.e. the infinite sum has not even an approximative result. But anyway we know that an infinite sum cannot have an equivalent in physical reality. So it's only consequent to consider finite partial sums
$$
\gamma_{2n}(x):=\sum_{m=0}^{n}Q0P\left(2m,0,\frac{1+\sqrt{1-x^2}}{2}\right)
$$
of (\ref{eqgammaprob}). $\gamma_\infty (x)=\gamma(x)$, additionally for every (finite) integer $n$ now also
\begin{equation}
\label{eqdefQ0partialsum}
\gamma_{2n}(1)=\gamma_n(-1)=\sum_{m=0}^{n}Q0P\left(2m,0,\frac{1}{2}\right)=
\sum_{m=0}^{n}Q0(2m,0)
\end{equation}
exists. It is not difficult to find a closed form for it. From
\begin{eqnarray*}
&&(2n-1) Q0(2n-2,0)+Q0(2n,0)=(2n-1)\frac{(2n-2)!}{2^{2n-2} \ (n-1)!^2}+\frac{(2n)!}{2^{2n} \ n!^2}\\
&=&\frac{2n \ (2n)!}{(2n)^2 \ 2^{2n-2} \  (n-1)!^2}+\frac{(2n)!}{2^{2n} \  n!^2}
=\frac{2n \ (2n)!}{n^2 \ 2^{2n} \ (n-1)!^2}+\frac{(2n)!}{2^{2n} \  n!^2}\\
&=&\frac{2n \ (2n)!}{2^{2n} \ n!^2}+\frac{(2n)!}{2^{2n} \ n!^2}=
(2n+1)\frac{(2n)!}{2^{2n} \ n!^2}\\
&=&(2n+1) Q0(2n,0)
\end{eqnarray*}
follows by induction 
$$
\sum_{m=0}^{n}Q0(2m,0)=(2n+1)\ Q0(2n,0)
$$
and with  (\ref{eqdefQ0partialsum})
\begin{eqnarray}
\gamma_{2n}(1)&=&(2n+1)\ Q0(2n,0) =(2n+1)\binom{2n}{n} \left( \frac{1}{2} \right)^{2n}\notag\\
&=&(2n+1) \ \frac{(2n)!}{2^{2n} \ n!^2}
\label{eqresultQ0partialsum}
\end{eqnarray}
\label{ChapterStirlingInCaseOfLargeN}
In case of large n we can use the Stirling formula $n!\approx n^n e^{-n} \sqrt{2 \pi n} $ and obtain $Q0(2n,0)\approx \frac{1}{\sqrt{\pi n}}$ and $\gamma_{2n}(1)\approx \sqrt{\frac{4n}{\pi}}$. So we have gotten a closed form for the sum (\ref{eqdefQ0partialsum}) of the return probabilities (the sum of the underlined values in table \ref{tableQ0} on page \pageref{tableQ0}). The results are finite even in case of $x=1$ (or $v=c$), because we assumed an only finite number $2n$ of steps. Obviously this assumption is adequate for all natural processes with finite duration.\\

\paragraph{Comment}
\wwnegeinr
The model of a one-dimensional random walk has only limited validity. Extensive considerations should take into account interactions between different reference systems and changes of the own reference system. Up to now we don't know enough about the exact ways of information between different reference systems\footnote{Example: Squared values like $\gamma_{2n}(1)^2\approx \frac{4 n}{\pi}$ or $\zeta_{2n}(1)^2 \approx \frac{1}{\pi n}$ (cf. (\ref{eqZeta1Closed}) in the appendix) can appear because of \emph{bi}directional information exchange during observation. The familiar macroscopic geometrical appearance isn't a primary thing, it's only a consequence of a discrete law \cite{or1}. The above considerations suggest an information theoretical approach to this law.} and about the long-term relation of their proper time. Further research is necessary, also combinatorial and graph theoretical research. The appendix demonstrates an example for possible connections of multiple random walks.

\section{Appendix}
We now introduce absorbing barriers, which are drains and can be sources of new random walks with steps in another orthogonal direction. Then we show, that in case of an absorbing barrier in the origin after start of the walk (and otherwise under the same the basic conditions as in theorem \ref{theoremgamma}) the probability of non-absorption is equivalent to $1/\gamma(x)=\sqrt{1-x^2}$. At last we investigate finite symmetric random walks with absorbing barrier.

\subsection{Absorbing barriers}
\label{ChapterAbsorbingBarriers}
A Bernoulli random walk can have absorbing barriers. If there is an absorbing barrier at point $a$ and the walking particle reaches it, the particle is absorbed. So point $a$ is only a drain, but no (direct) source for further walks within the same dimension\footnote{It can be source of a walk in another dimension.}. We can get the resulting probability distribution by subtraction of a "shifted" distribution from (\ref{eqdefq0p}): Let's assume an absorbing barrier at $a>0$. We define
\begin{equation}
\label{eqdefpa}
P_a(n,k,p):=Q0P(n,k,p)-\left(\frac{p}{1-p}\right)^a Q0P(n,k-2a,p)
\end{equation}
from which follows
\begin{equation}
\label{eqabsorbedwalklaw}
P_a(n + 1, k, p) = p \ P_a(n, k - 1, p) + (1 - p) \  P_a(n, k + 1, p) \ ,
\end{equation}
i.e. the inductive law (\ref{eqwalklaw}) of a Bernoulli random walk holds. Additionally the boundary condition $P_a(n,a,p)=0$ is fulfilled, so that point $a$ is only drain, but not source\footnote{In literature at point $a$ often the sum of absorption probabilities is listened. Here this special treatment is not done, so that law (\ref{eqabsorbedwalklaw}) is valid.}. Therefore $P_a(n,k,p)$ represents for all values $-n \le k \le a$ within reach the probability, that the particle passes point $k$ and continues moving. For the particle starting at $k<a$ the points $k>a$ are not within reach\footnote{$P_a(n,k,p)$ is negative there. In case of a simultaneous walk of two particles with starting points $0$ and $2a$, in which the particle starting at $2a$ is the annihilating counterpart of the other starting at $0$, for $k>a$ the absolute value $|P_a(n,k,p)|$ can be interpreted as probability, that the annihilating counterpart passes point $k$, if both particles make simultaneously steps in opposite directions. If this is not guaranteed, there is a chance, that a particle passes the barrier (like in the tunnel effect).}.

\subsubsection{Random walk with delayed absorbing barrier at $k=0$}
\label{WalkWithAbsorbingBarrierIn0}
The starting coordinate $k=0$ plays a special role and it is reasonable to assume an absorbing barrier there, also because of symmetry. But if this barrier is active from the beginning on, the particle is absorbed at once so that the walk cannot begin and (\ref{eqdefpa}) has the meaningless result $P_0(n,k,p)=Q0P(n,k,p)-Q0P(n,k,p)=0$. However, if there is absorption at $k=0$ \emph{after} the walk already has started, we get a distribution which is worth further consideration. So let's assume a \emph{delayed} absorbing barrier at $k=0$ which is activated \emph{after} completion of the first step of the walk. The resulting probability distribution is given by the absolute values of
\begin{equation}
\label{eqdefq1p}
Q1P(n,k,p):=(1-p)\ Q0P(n-1,k+1,p)-p\ Q0P(n-1,k-1,p)
\end{equation}
which is a modification\footnote{$Q1P(n,k,p)=(1-p)\ P_1(n-1, k+1, p)$; an absorbing barrier at $k=1$ is within reach and therefore active only from the second step on. $Q1P(n,k,p)$ is "centered" in this barrier.} of (\ref{eqdefpa}). $Q1P$ fulfills the boundary conditions
$$
|Q1P(0,0,p)|=1, \ Q1P(2n,0,p)=0 \text{ for } n\ge 1 \ 
$$
and the same inductive law as $Q0P$ in (\ref{eqwalklaw}). For $n\ge 1$ a more compact form of $Q1P(n,k,p)$ is
\begin{equation}
\label{eqq1pcompact}
Q1P(n, k, p)= \frac{- k}{n} Q0P(n, k, p)
\end{equation}

because of

\small
\begin{eqnarray*}
&&Q1P(n, k, p)=\ (1 - p) \ Q0P(n - 1, k + 1, p) - p\ Q0P(n - 1, k - 1, p)\\
&=&\frac{p^{(n + k)/2} \ (1 - p)^{(n - k)/2} \ (n - 1)!}{\left(\frac{n+k}{2}\right)! \left(\frac{n - k}{2}-1\right)!}
- \frac{p^{(n + k)/2} \ (1 - p)^{(n - k)/2} \ (n - 1)!}{\left(\frac{n+k}{2}-1\right)! \left(\frac{n - k}{2}\right)!}\\
&=& \left(\frac{n - k}{2 n}\right) \frac{p^{(n + k)/2} \ (1 - p)^{(n - k)/2} \ n!}{\left(\frac{n+k}{2}\right)! \left(\frac{n - k}{2}\right)!}
-  \left(\frac{n + k}{2 n}\right) \frac{p^{(n + k)/2} \ (1 - p)^{(n - k)/2} \ n!}{\left(\frac{n+k}{2}\right)! \left(\frac{n - k}{2}\right)!}\\
&=& \left(\frac{- k}{n}\right) \frac{p^{(n+k)/2} \ (1 - p)^{(n - k)/2} \ n!}{\left(\frac{n+k}{2}\right)! \left(\frac{n - k}{2}\right)!}
= \frac{- k}{n} \ Q0P(n, k, p) \ .
\end{eqnarray*}
\normalsize

\subsubsection{Past differences}
\label{ChapterPastDifferences}
Equation (\ref{eqdefq1p}) has similarity to a finite difference along $k$. It represents the probability difference of the two ways coming from past. Therefore we shall call the accompanying operator \emph{past difference} and use the symbol $\dpast$ for it. If $\psi$ is a function of the variables $n,k,p$, defined at least at $(n-1,k+1,p)$ and $(n-1,k-1,p)$, its past difference is
\begin{equation}
\label{eqdefdpast}
\dpast \psi(n,k,p)=(1-p)\ \psi(n-1,k+1,p)-p\ \psi(n-1,k-1,p) \ .
\end{equation}
Similarly to usual finite differences we can form higher-order past differences, for example the second-order past difference 
\begin{eqnarray}
&&Q2P(n,k,p):=\dpast^2 Q0P(n,k,p)=\dpast\dpast Q0P(n,k,p)\notag\\
&=&(1-p)\ \dpast Q0P(n-1,k+1,p)-p\ \dpast Q0P(n-1,k-1,p)\notag\\
\label{eqdefq2p}
&=&(1 - p)^2 \ Q0P(n - 2, k + 2, p) + p^2\ Q0P(n - 2, k - 2, p)\\
&& - 2p(1 - p)\ Q0P(n - 2, k, p) \ .\notag 
\end{eqnarray}

For $n \ge 2$ we obtain
\footnotesize
\begin{eqnarray*}
&&\dpast^2 Q0P(n,k,p)=\dpast\dpast Q0P(n,k,p)=\dpast Q1P(n,k,p)\\
&=&\dpast\left(\frac{-k}{n} \ Q0P(n,k,p)\right)\\
&=&(1-p)\left(\frac{-k-1}{n-1}\right) Q0P(n-1,k+1,p)-p\ \left(\frac{1-k}{n-1}\right) Q0P(n-1,k-1,p)\\
&=&\left(\frac{-k-1}{n-1}\right)\left((1-p) \ Q0P(n-1,k+1,p)-p\ Q0P(n-1,k-1,p)\right)\\
&&-p \left(\frac{2}{n-1}\right) Q0P(n-1,k-1,p) \\
&=&\left(\frac{-k-1}{n-1}\right) Q1P(n,k,p)
-\left(\frac{2}{n-1}\right) \left(\frac{n+k}{2n}\right) Q0P(n,k,p) \\
&=&\left(\frac{k(k+1)}{n(n-1)}\right) Q0P(n,k,p)- \left(\frac{n+k}{n(n-1)}\right) Q0P(n,k,p) \\
&=&\frac{k^2-n}{n(n-1)} \ Q0P(n,k,p)\ .
\end{eqnarray*}
\normalsize
The \emph{central} second-order past differences
\begin{equation}
\label{eqcentral2pd}
Q2P(2n,0,p)=\frac{-1}{2n-1}Q0P(2n,0,p)=\frac{-1}{2n-1}\binom{2n}{n} (1-p)^n \ p^n\ 
\end{equation}
have a special meaning: Because of
\begin{eqnarray*}
|Q2P(2n,0,p)|&=&|\dpast Q1P(2n,0,p)|\\
&=&|(1-p)\ Q1P(2n-1,1,p)-p\ Q1P(2n-1,-1,p)|\\
&=&(1-p)\ |Q1P(2n-1,1,p)|+p\ |Q1P(2n-1,-1,p)|
\end{eqnarray*}
for $n \ge 1$ the absolute values
\begin{equation}
\label{eqQ2PAbsoprtionProbability}
|Q2P(2n,0,p)|=\frac{Q0P(2n,0,p)}{2n-1}
\end{equation}
correspond to the probability of absorption after the $2n$-th step of the random walk specified in chapter \ref{WalkWithAbsorbingBarrierIn0}.\\

It is worth\footnote{In important physical equations (e.g. Schrödinger equation) the second derivative along location is related to the first derivative along time.} mentioning that the second order past difference (along $k$) is equivalent to a weighted first order difference along $n$:
\begin{equation}
\label{eqQ2PasTimeDiff}
Q2P(n,k,p) = Q0P(n, k, p) - 4 p (1 - p)\ Q0P(n - 2, k, p) \ .
\end{equation}

This follows from (\ref{eqdefq2p}) and
\begin{eqnarray*}
Q0P(n, k, p)&=&(1 - p)^2 \ Q0P(n - 2, k + 2, p) + p^2 \ Q0P(n - 2, k - 2, p)\\
&& + 2p(1 - p)\ Q0P(n - 2, k, p) \ .\\
\end{eqnarray*}

\subsection{The power series of $1/\gamma(x)=\sqrt{1-x^2}$}
Just like in chapter \ref{Chapterpowseriesgamma} we now look at the power series of
\begin {equation}
\zeta: \ [-1,1 ]\ \to \wwwr, \ \ \zeta(x)=\sqrt{1-x^2} \ ;\notag
\end {equation}
$\zeta(x)=\frac{1}{\gamma(x)}$ for $|x|<1$ and $\zeta(-1)=\zeta(1)=0$\ .
${\sqrt{1+z}}=f_{1/2}(z)$, so analogously to (\ref{eqherlgamma}) we get

\begin{eqnarray*}
\sqrt{1+z}
&=&\sum_{l=0}^{\infty}\binom{\frac{1}{2}}{l}z^{l} \\
&=&1+ \frac{\frac{1}{2}}{1} z^1+ \frac{\frac{1}{2} \cdot - \frac{1}{2}}{1 \cdot 2} z^2
+ \frac{\frac{1}{2} \cdot - \frac{1}{2} \cdot - \frac{3}{2}}{1 \cdot 2 \cdot 3} z^3
+ \frac{\frac{1}{2} \cdot - \frac{1}{2} \cdot - \frac{3}{2} \cdot - \frac{5}{2}}{1 \cdot 2 \cdot 3 \cdot 4} z^4+... \\
&=&1+ \frac{1}{2^1 \cdot 1!} z^1- \frac{1 \cdot 1}{2^2 \cdot 2!} z^2
+ \frac{1 \cdot 1 \cdot 3}{2^3 \cdot 3!} z^3
- \frac{1 \cdot 1 \cdot 3 \cdot 5}{2^4 \cdot 4!} z^4+... \\
&=&1-\sum_{l=1}^{\infty}\frac{1}{2l-1} \ \binom{2l}{l} \left( \frac{-z}{4} \right)^{l} \ 
\end{eqnarray*}
from which follows
\begin{equation}
\zeta(x)= \sqrt{1-x^2}=1-\sum_{l=1}^{\infty}\frac{1}{2l-1} \ \binom{2l}{l} \left( \frac{x}{2} \right)^{2l} \ .\notag
\end{equation}
So in case of $4p(1-p)=x^2$ we obtain with (\ref{Q0PsubstitutedP}), (\ref{eqp1p}) and (\ref{eqcentral2pd})
\begin{eqnarray}
&&\zeta(x)=\sqrt{1-x^2}=1-\sum_{n=1}^{\infty}\ \frac{1}{2n-1} Q0P(2n,0, p) \notag\\
&=&1+\sum_{n=1}^{\infty} \ Q2P(2n,0, p)=1-\sum_{n=1}^{\infty} \ |Q2P(2n,0,p)|\ .\notag
\end{eqnarray}

Because $|Q2P(2n,0,p)|$ is the probability of absorption after the $2n$-th step, $\sum_{n=1}^{\infty} \ |Q2P(2n,0,p)|$ is the total probability of absorption. Therefore we conclude:

\subsubsection{Theorem: $\sqrt{1-x^2}$ as probability of non-return (of "escape")}
\label{theoremZetaPropNonabs}
If a particle makes a Bernoulli random walk, in which each step is directed from point $k$ to $k+1$ with probability $p$, from point $k$ to $k-1$ with probability $1-p$ and $4p(1-p)=x^2$ and the particle is absorbed if it returns to the starting point, the probability of non-absorption (of "escape") is $\zeta(x)=\sqrt{1-x^2}$.

\subparagraph{Remark.}
Also more concrete formulations of this theorem are possible. Due to experimental results we know, that the energy of a photon can be distributed. If $E$ is the energy of the photon, its frequency $\nu$ is given by $\nu=\frac{E}{h}$, in which $h$ is Planck's constant ($h\approx 6.626 \cdot 10^{-34}\ Js$). At this a reduction of the photon's frequency is equivalent to a dilation of its time period. So we can state:

\subsubsection{Theorem: Energy of a received photon as non-returning (escaping) part of its initial energy}
Let $\gamma(x)=\frac{1}{\sqrt{1-x^2}}$ represent the (approximative) time dilation factor of reference system A relative to reference system B as in theorem \ref{theoremgamma}. If a photon is emitted in B with energy $E_e=h\nu_e$ and absorbed in A, the maximal absorption energy $E_a=h\nu_a$ in system A is given by $E_a=E_e \sqrt{1-x^2}$. So the quotient\footnote{The part of the photon's energy which can escape and arrive in A in comparison to initial energy of the photon} $\frac{E_a}{E_e}=\frac{\nu_a}{\nu_e}$ is (approximatively) equivalent to the probability\footnote{The expectation value of the frequency of non-returning (escaping) walks in comparison to the total frequency or total number of walks}, that there is no return to the starting point during a Bernoulli random walk, in which each step is directed from point $k$ to $k+1$ with probability $p$, from point $k$ to $k-1$ with probability $1-p$ and $4p(1-p)=x^2 $.

\subsection{Case $x=1$ resp. $v=c$ with absorbing barrier}
\subsubsection{Symmetry}
In case of $x=1$ or $v=c$ also the chapter \ref{WalkWithAbsorbingBarrierIn0} described random walk with absorbing barrier becomes symmetric, because the (after the first step active) barrier is located in the starting point $k=0$ and $p=1-p=\frac{1}{2}$ with (\ref{eqp1p}). The probability, that after the $n$-th step point $k$ is reached and the walk continues, is given by the absolute value of
\begin{equation}
Q1(n,k):=Q1P\left(n,k,\frac{1}{2}\right)\ .\notag
\end{equation}
\begin{table}[h]
$ $\\
\setlength{\tabcolsep}{0.14cm}
\begin{tabular}{rrrrrrrrrrrrrrrrrr}
$n$ &$k\to$ & $-6$ & $-5$ & $-4$ & $-3$ & $-2$ & $-1$ & 
$\ \ 0$ & $\ \ 1$ & $\ \ 2$ & $\ \ 3$ & $\ \ 4$ & $\ \ 5$ & $\ \ 6$ & &\\ 

$\downarrow$ &$  $ & $  $ & $  $ & $  $ & $  $ & $  $ & $  $ & 
$  $ &$  $ & $  $ & $  $ & $  $ & $  $ & $  $ & $  $ & $$\\

$ 1$ &$  $ & $  $ & $  $ & $  $ & $  $ & $  $ & $ \underline{1}$ & 
$  $ &$-1$ & $  $ & $  $ & $  $ & $  $ & $  $ & $  $ & $\ \cdot 2^{-1}$\\

$ 2$ &$  $ & $  $ & $  $ & $  $ & $  $ & $ 1$ & $  $ & 
$ 0$ &$  $ & $-1$ & $  $ & $  $ & $  $ & $  $ & $  $ & $\ \cdot 2^{-2}$\\

$ 3$ &$  $ & $  $ & $  $ & $  $ & $ 1$ & $  $ & $ \underline{1}$ & 
$  $ &$-1$ & $  $ & $-1$ & $  $ & $  $ & $  $ & $  $ & $\ \cdot 2^{-3}$\\

$ 4$ &$  $ & $  $ & $  $ & $ 1$ & $  $ & $ 2$ & $  $ & 
$ 0$ &$  $ & $-2$ & $  $ & $-1$ & $  $ & $  $ & $  $ & $\ \cdot 2^{-4}$\\

$ 5$ &$  $ & $  $ & $ 1$ & $  $ & $ 3$ & $  $ & $ \underline{2}$ & 
$  $ &$-2$ & $  $ & $-3$ & $  $ & $-1$ & $  $ & $  $ & $\ \cdot 2^{-5}$\\

$ 6$ &$  $ & $ 1$ & $  $ & $ 4$ & $  $ & $ 5$ & $  $ & 
$ 0$ &$  $ & $-5$ & $  $ & $-4$ & $  $ & $-1$ & $  $ & $\ \cdot 2^{-6}$\\

$...$ &$  $ & $  $ & $  $ & $  $ & $  $ & $  $ & $  $ & 
$  $ &$  $ & $  $ & $  $ & $  $ & $  $ & $  $ & $  $ & $$
\end{tabular}
\caption{The first values of $Q1(n,k)=Q1(n,k,\frac{1}{2})$. $|Q1(n,k)|$ is the probability, that after the $n$-th step point $k$ is reached and the walk continues, therefore ${Q1(2n,0)=0}$ in the absorbing barrier. The underlined values ${Q1(2n-1,-1)=|Q1(2n-1,1)-Q1(2n-1,-1)|/2=|Q2P(2n,0,\frac{1}{2})|}$ are the probabilities of absorption after the $2n$-th step. It is visible, that the numbers result from addition of two symmetric binomial distributions with opposite sign, one starting at $(n,k)=(1,-1)$, the other starting at $(n,k)=(1,1)$, so that at $k=0$ annihilation occurs.}
\label{tableQ1}
\end{table}

\subsubsection{Finite random walk}
With \ref{theoremZetaPropNonabs} in case of $x=1$ the probability of absorption (or return to the starting point) is $1$ if the number of steps in the walk has no upper limit. Because in physical reality within finite time only a finite number of steps are possible we consider finite partial sums
$$
\zeta_{2n}(x):=1+\sum_{m=1}^{n} \ Q2P\left(2m,0, \frac{1+\sqrt{1-x^2}}{2}\right)\ 
$$
of the power series of $\zeta(x)$. Similarly as in chapter \ref{ChapterGammaPartialsum} for $\gamma_{2n}(1)$ we can find a closed form for $\zeta_{2n}(1)$. For $n>0$ we get with (\ref{eqQ2PasTimeDiff})
$$
Q0P\left(n - 2, 0, \frac{1}{2}\right)+ Q2P\left(n, 0, \frac{1}{2}\right) = Q0P\left(n, 0, \frac{1}{2}\right)
$$
so that with $Q0P\left(0, 0, \frac{1}{2}\right)=1$ by induction follows
$$
\zeta_{2n}(1)=1+\sum_{m=1}^{n} \ Q2P\left(2m,0, \frac{1}{2}\right)= Q0P\left(2n,0, \frac{1}{2}\right)
=\frac{(2n)!}{2^{2n}(n!)^2} .
$$
In case of large $n$ we can use the Stirling formula and obtain
\begin{equation}
\label{eqZeta1Closed}
\zeta_{2n}(1)\approx \frac{1}{\sqrt{\pi n}}\ .
\end{equation}
$1-\zeta_{2n}(1)$ is the probability of absorption in case of $x=1$ or $p=1-p=\frac{1}{2}$ when making at most $2n$ steps. The probability of absorption (\ref{eqQ2PAbsoprtionProbability}) after the $2n$-th step is given by the negative second-order past difference (along $k$)
\begin{eqnarray*}
-Q2P\left(2n,0,\frac{1}{2}\right)&=&-\dpast^2 Q0P\left(2n,0,\frac{1}{2}\right)=\frac{1}{2n-1}Q0P\left(2n,0,\frac{1}{2}\right)\\
&\approx& \frac{1}{\sqrt{4\pi n^3}}\ 
\end{eqnarray*}
and because of the Schrödinger equation it is remarkable, that with (\ref{eqQ2PasTimeDiff}) this is equivalent to the negative (first-order) finite difference along $n$:
$$
-Q2P\left(2n,0,\frac{1}{2}\right) = Q0P\left(2n-2,0, \frac{1}{2}\right)-Q0P\left(2n, 0, \frac{1}{2}\right) \ .
$$\\ 

We have seen, that the in chapter \ref{ChapterPastDifferences} defined "discrete differentiation" leads to a probability distribution with absorbing barrier. Separation (and distinction) of the ways on both sides of the barrier is connected with this. We should recall, that in physical experiments (e.g. double slit experiment) such separation also is connected with absorption - and emission - of photons at systems with rest mass.

\newpage


\begin{thebibliography}{\hspace{3em}}

\bibitem{barber}
M.N. Barber, B.W. Ninham,
\emph{Random and restricted walks, theory and applications},
New York: Gordon \& Breach, 1970.

\bibitem{feller}
W. Feller,
\emph{An introduction to probability theory and its applications},
Vol. 1-2, New York: Wiley, 1957-1971.

\bibitem{fi}
A. Fine,
\emph{Theories of probabilities, an examination of foundations},
New York: Academic Press, 1973.

\bibitem{gnedenko}
B.V. Gnedenko,
\emph{The theory of probability},
New York: Chelsea, 1963.

\bibitem{go}
S. Gottwald,
\emph{Meyers kleine Enzyklopädie Mathematik, 14. Auflage},
Mannheim, Leipzig, Wien, Zürich: Meyers Lexikonverlag, 1995.

\bibitem{gut}
A. Gut,
\emph{Stopped random walks. Limit theorems and applications},
New York: Springer, 1988.

\bibitem{kac}
M. Kac,
\emph{Statistical independence in probability, analysis and number theory},
Buffalo, NY: Math. Assoc. Amer., 1963.

\bibitem{kr}
U. Krengel,
\emph{Einführung in die Wahrscheinlichkeitstheorie und Statistik},
3. erw. Aufl., Braunschweig: Vieweg, 1991.

\bibitem{khrennikov}
A. Khrennikov, Y. Volovich,
\emph{Discrete Time Leads to Quantum-Like Interference of Deterministic Particles},
quant-ph/0203009.

\bibitem{kn}
K. Knopp,
\emph{Theorie und Anwendung der unendlichen Reihen},
Berlin, Heidelberg, New York: Springer, 1964.

\bibitem{mac}
C. MacLaurin,
\emph{A treatise of fluxions},
Vol. 1-2, Edinburgh, 1742.

\bibitem{mar}
A.I. Markushevich,
\emph{Theory of functions of a complex variable},
Vol. 1, New York: Chelsea, 1977.

\bibitem{mis}
C.W. Misner, K.S. Thorne, J.A. Wheeler,
\emph{Gravitation},
New York: W.H. Freeman and Company, 1973.

\bibitem{or}
W. Orthuber,
\emph{The Recombination Principle: Mathematics of decision and perception}  (init. 2000),\ \ 
 http://www.orthuber.com/ 

\bibitem{or1}
W. Orthuber,
\emph{To the finite information content of the physically existing reality}, quant-ph/0108121.

\bibitem{spi}
F. Spitzer,
\emph{Principles of random walk, 2nd ed.},
New York: Springer-Verlag, 1976.

\bibitem{tr}
A.S. Troelstra,
\emph{Choice Sequences; A Chapter of Intuitionistic Mathematics},
Oxford: Clarendon Press, 1977.\\ \\


Special references to the binomial distribution, its extensions and modifications:


\bibitem{firstbinref}
M. Bauer, C. Godreche, J.M. Luck,
\emph{Statistics of persistent events in the binomial random walk: Will the drunken sailor hit the sober man?},
cond-mat/9905252.

\bibitem{bauer1}
W. Bauer, S. Pratt,
\emph{Size Matters: Origin of Binomial Scaling in Nuclear Fragmentation Experiments},
nucl-th/9808068.

\bibitem{beenakker}
C.W.J. Beenakker, H. Schomerus,
\emph{Counting statistics of photons produced by electronic shot noise},
cond-mat/0008413.

\bibitem{berkovich}
A. Berkovich,
\emph{Fermionic counting of RSOS-states and Virasoro character formulas for the unitary minimal series $M(\nu ,\nu +1)$. Exact results},
hep-th/9403073.

\bibitem{bottini} S. Bottini,
\emph{New Polynomial Law of Hadron Mass},
hep-ph/9810405.

\bibitem{brooks}
MiniMax Collaboration: T. C. Brooks et al.,
\emph{A Search for Disoriented Chiral Condensate at the Fermilab Tevatron},
hep-ex/9906026.

\bibitem{fu}
Hong-Chen Fu, Ryu Sasaki,
\emph{Negative Binomial States of Quantized Radiation Fields},
quant-ph/9610024.

\bibitem{fu1}
Hong-Chen Fu, Ryu Sasaki,
\emph{Probability Distributions and Coherent States of $B_r$, $C_r$ and $D_r$ Algebras},
hep-th/9706034.

\bibitem{giovannini}
A. Giovannini,
\emph{Clan concept in multiparticle dynamics and the NB ``enigma''},
hep-ph/9611408.

\bibitem{giovannini1}
A. Giovannini, S. Lupia, R. Ugoccioni,
\emph{The negative binomial distribution in quark jets with fixed flavour},
hep-ph/9609306.

\bibitem{giovannini2}
A. Giovannini, S. Lupia, R. Ugoccioni,
\emph{Common origin of the shoulder structure and of the oscillations of moments in multiplicity distributions in e(+)e(-) annihilations},
hep-ph/9609406.

\bibitem{hara}
P. O'Hara,
\emph{Clebsch-Gordan coefficients and the binomial distribution},
quant-ph/0112096.

\bibitem{hegyi}
S. Hegyi,
\emph{Multiplicity Distributions in Strong Interactions: A Generalized Negative Binomial Model},
hep-ph/9608346.

\bibitem{hegyi1}
S. Hegyi,
\emph{H-function extension of the NBD in the light of experimental data},
hep-ph/9707322.

\bibitem{hegyi2}
S. Hegyi,
\emph{H-function extension of the NBD: further applications},
hep-ph/9708241.

\bibitem{lee}
H. W. Lee, L. S. Levitov,
\emph{Orthogonality catastrophe in a mesoscopic conductor due to a time-dependent flux},
cond-mat/9312013.

\bibitem{lee1}
H. W. Lee, L. S. Levitov,
\emph{Estimate of Minimal Noise in a Quantum Conductor},
cond-mat/9507011.

\bibitem{levitov}
L. S. Levitov, H.-W. Lee, G. B. Lesovik,
\emph{Electron Counting Statistics and Coherent States of Electric Current},
cond-mat/9607137.

\bibitem{levitov1}
L.S.Levitov, G.B.Lesovik,
\emph{Quantum Measurement in Electric Circuit},
cond-mat/9401004.

\bibitem{martinis}
M. Martinis, V. Mikuta-Martinis,
\emph{Intensity-dependent pion-nucleon coupling in multipion production processes},
nucl-th/9901031.

\bibitem{matinyan}
S.G. Matinyan, E.B. Prokhorenko,
\emph{Branching Processes and Multi-Particle Production},
hep-ph/9305226.

\bibitem{moretto}
L.G. Moretto, L. Beaulieu, L. Phair, G.J. Wozniak,
\emph{Statistical Exploration of Fragmentation Phase Space Source Sizes in Nuclear Multifragmentation},
nucl-ex/9812010.

\bibitem{moretto1}
 L.G. Moretto, R. Ghetti, K.X. Jing, L. Phair, K. Tso, G.J. Wozniak,
\emph{The Role of Phase Space in Complex Fragment Emission from Low to Intermediate Energies},
nucl-ex/9607014.

\bibitem{nakajima}
N. Nakajima, M. Biyajima, N. Suzuki,
\emph{Analysis of Cumulant Moments in High Energy Hadron-Hadron Collisions by Truncated Multiplicity Distributions},
hep-ph/9606369.

\bibitem{puri}
R.R. Puri, S. Arun Kumar, R.K. Bullough,
\emph{Stroboscopic theory of atomic statistics in the micromaser},
quant-ph/9910103.

\bibitem{tchikilev}
O. G. Tchikilev,
\emph{Phenomenological Parametrization of the Charged Particle Multiplicity Distributions in Restricted Rapidity Intervals in e+e- Annihilation into Hadrons and e+p Scattering at HERA},
hep-ph/9612214.

\bibitem{tchikilev1}
O.G. Tchikilev,
\emph{Multiplicity distributions in e+e- annihilation into hadrons and pure birth branching processes},
hep-ph/9912237.

\bibitem{ugoccioni}
R. Ugoccioni, A. Giovannini,
\emph{Oscillations of Moments and Structure of Multiplicity Distributions in e+e- Annihilation},
hep-ph/9710360.

\bibitem{wang}
Xiao-Guang Wang, Barry C. Sanders, Shao-hua Pan,
\emph{Entangled SU(2) and SU(1,1) coherent states},
quant-ph/0001073.

\bibitem{wang1}
Xiao-Guang Wang, Hong-Chen Fu,
\emph{Negative Binomial States of the Radiation Field and their Excitations are Nonlinear Coherent States},
quant-ph/9903013.

\bibitem{lastbinref}
Xiao-Guang Wang, Shao-Hua Pan, Guo-Zhen Yang,
\emph{Non-classical properties and algebraic characteristics of negative binomial states in quantized radiation fields},
quant-ph/9904027.

\end{thebibliography}
\end{document}